\documentclass[letterpaper,twocolumn,prl, showpacs]{revtex4}
\usepackage{graphicx, amsmath,amssymb}
\usepackage[colorlinks=true, linkcolor=blue, citecolor=blue, urlcolor=blue]{hyperref}

\begin{document}

\title{Exploring Energy-Time Entanglement Using Geometric Phase }
\author{Anand Kumar Jha, Mehul Malik, and Robert W. Boyd}
\affiliation{The Institute of Optics, University of Rochester,
Rochester, New York 14627, USA}
\date{\today}

\begin{abstract}

Using the signal and idler photons produced by parametric
downconversion, we report an experimental observation of a
violation of the Bell inequality for energy and time based purely
on the geometric phases of the signal and idler photons. We thus
show that energy-time entanglement between the signal and idler
photons can be explored by means of their geometric phases. These
results may have important practical implications for quantum
information science by providing an additional means by which
entanglement can be manipulated.\\

\noindent \textbf{Published in: Phys. Rev. Lett. 101, 180405 (2008) \href{https://doi.org/10.1103/PhysRevLett.101.180405}{10.1103/PhysRevLett.101.180405}}

\end{abstract}


\maketitle

Geometric phase, or Berry's phase, is the phase acquired by a
system when it is transported around a closed circuit in an
abstract space \cite{berry-phase}. The manifestation of this phase
in polarization optics is also known as Pancharatnam phase, which
is the phase acquired by a photon field when its polarization is
taken through a closed circuit on the Poincar\'{e} sphere
\cite{pancharatnam, ramasheshan, berry-connection}. Pancharatnam
phase has been observed both at high light levels \cite{bhandari,
mandel-geometric, chiao-geometric}  and at a single photon level
\cite{kwiat-geometric}. Effects of Pancharatnam phase in
two-photon interference, using the signal and idler photons
produced by parametric downconversion (PDC), have also been
studied in many different situations \cite{brendel-geometric,
shih-geometric, induced-geometric, mehul-geometric, Sjöqvist}.

Bell's inequality \cite{bell-inequality} was derived in order to
show that any local hidden variable interpretation  \cite{bohm} of
quantum mechanics is incompatible with the statistical predictions
of quantum mechanics. Clauser, Horne, Shimony and Holt (CHSH)
generalized Bell's inequality so that it could be applied to
realizable experiments \cite{chsh}. Since then, using the signal
and idler photons produced by PDC, violations of the generalized
forms of Bell's inequality have been observed for various degrees
of freedom including polarization \cite{bell-polarization1,
bell-polarization2}, phase and momentum
\cite{bell-linear-momentum}, and energy and time
\cite{bell-time-energy1, bell-time-energy2, bell-time-energy3}. In
recent years, using hyperentangled states, even simultaneous
violations of Bell inequalities for more than one degrees of
freedom have been reported \cite{hyper-polar+time-energy,
hyper-polar+momentum1, hyper-polar+momentum2,
hyper-polar+spatial+time-energy}. In addition to proving the
impossibility of local hidden variable interpretations of quantum
mechanics, a violation of Bell's inequality, based on a certain
degree of freedom, verifies entanglement and guarantees that it
can be exploited through that particular degree of freedom.

Bell inequality for energy and time was suggested by J. D. Franson
using an experimental scheme commonly known as the Franson
interferometer \cite{bell-time-energy1}. Franson's scheme for
violating a Bell inequality requires changing the phases of the
signal and idler photons in one of the interfering alternatives.
In all the experimental realizations of Franson's scheme so far,
the phases of the signal and idler photons have been changed by
adjusting their dynamic phases, i.e., by adjusting their optical
path lengths \cite{bell-time-energy2, bell-time-energy3}.
Therefore, all these previous violations can be said to be the
dynamic phase-based violations of Bell inequality for energy and
time.

In this paper, we show that the Bell inequality for energy and
time can also be violated using geometric phases of the signal and
idler photons and that therefore the energy-time entanglement can
be explored using their geometric phases. Throughout this paper,
we use Ref. \cite{two-photon coherence} as our theoretical
framework.
\begin{figure}[b!]
\includegraphics{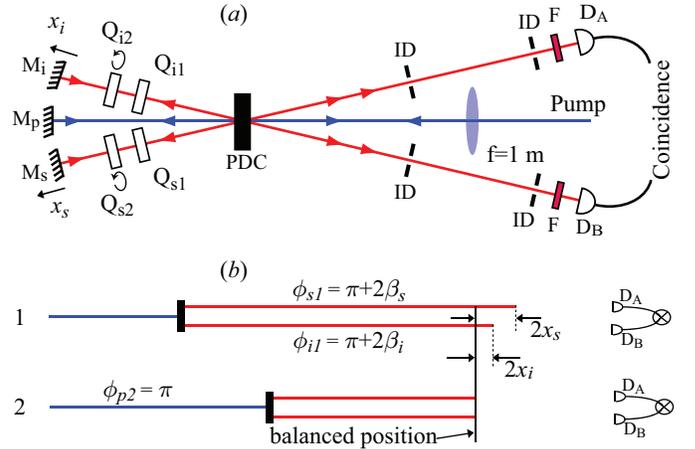}
  \caption{(color online). ($a$) Schematic of the experimental setup.  Q$_{\textrm{s1}}$ and Q$_{\textrm{i1}}$ are quarter-wave plates with their optic axes
oriented at $45^{\circ}$ from the horizontal polarization
direction. Q$_{\textrm{s2}}$ and Q$_{\textrm{i2}}$ are rotatable
quarter-wave plates with their optic axes oriented at angles
$135^{\circ}+\beta_{s}$ and $135^{\circ}+\beta_{i}$ from the
horizontal polarization direction respectively. F is an
interference filter with 10-nm bandwidth, centered at 727.6 nm; ID
is an iris diaphragm. The signal and idler photons are collected
into multimode fibers and detected using two avalanche
photodiodes. A weak lens (f=1 m) focuses the pump beam onto the
pump mirror ${\rm M_{p}}$. ($b$) Two-photon path diagrams
illustrating how the geometric phases $2\beta_s$ and $2\beta_i$
influence the two-photon interference.}\label{setup}
\end{figure}

Consider the double-pass setup \cite{frustrated two-photon} shown
in Fig.~\ref{setup}($a$). A cw Ar-ion laser operating at 363.8 nm
is used as a pump to produce degenerate type-I parametric
downconverson (PDC). The pump is vertically polarized and the
downconverted photons are both horizontally polarized. The
quarter-wave plates Q$_{\textrm{s1}}$ and Q$_{\textrm{i1}}$ are
arranged with their optic axes oriented at angle $45^{\circ}$ from
the horizontal polarization direction, while the rotatable
quarter-wave plates Q$_{\textrm{s2}}$ and Q$_{\textrm{i2}}$ are
arranged with their optic axes oriented at angles
$135^{\circ}+\beta_{s}$ and $135^{\circ}+\beta_{i}$ from the
horizontal polarization direction respectively. In this setup
there are two alternative pathways---represented by the two-photon
path diagrams in Fig.~\ref{setup}($b$)---by which the pump photon
gets downconverted and the downconverted photons get detected at
single-photon detectors $D_{\textrm{A}}$ and $D_{\textrm{B}}$. In
alternative 1, a pump photon gets downconverted in its forward
pass and the downconverted signal and idler photons reach the two
detectors $D_{\textrm{A}}$ and $D_{\textrm{B}}$ after passing
through the quarter-wave plates and getting reflected from the
signal (${\rm M_{s}}$) and idler (${\rm M_{i}}$) mirrors. In
alternative 2, a pump photon gets downconverted after getting
reflected from the pump mirror (${\rm M_{p}}$) and the
downconverted photons directly reach their respective detectors.

\begin{figure}[t!]
\includegraphics{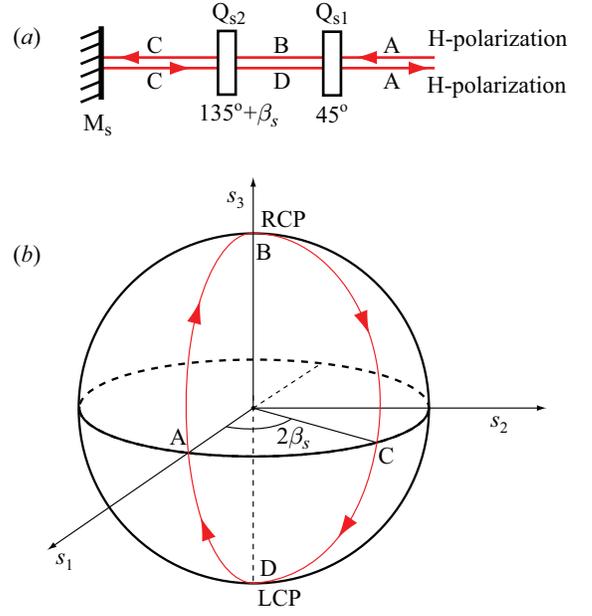}
  \caption{(color online). ($a$) The signal photon path ABCDA through the two quarter-wave plates Q$_{\textrm{s1}}$
  and Q$_{\textrm{s2}}$, in alternative 1. ($b$) Poincar\'{e}-sphere
representation of polarization states of the signal photon
corresponding to path ABCDA. A signal photon acquires a phase
2$\beta_{s}$ in going through path ABCDA, which is equal to half
of the solid angle subtended by the closed circuit ABCDA on the
Poincar\'{e} sphere. }\label{poincare}
\end{figure}

In alternative 1, a horizontally polarized signal photon passes
through the quarter-wave plates Q$_{\textrm{s1}}$ and
Q$_{\textrm{s2}}$, gets reflected back from the signal mirror
${\rm M_{s}}$ and retraces its path through the quarter-wave
plates Q$_{\textrm{s2}}$ and Q$_{\textrm{s1}}$. It becomes
horizontally polarized after completing the loop but in this
process it acquires a geometric phase equal to $2\beta_{s}$, as
illustrated in Fig.~\ref{poincare}. This phase is in addition to
the dynamic phase that the signal photon acquires. Similarly, the
idler photon acquires a geometric phase equal to $2\beta_{i}$ in
alternative 1.
In the balanced position of the setup in Fig.~\ref{setup}($a$),
the optical path lengths between the crystal and each of the three
mirrors are assumed to be equal, to about 15 cm. The displacements
of the signal and idler mirrors from the balanced position are
denoted by $x_{s}$ and $x_{i}$ respectively.

We recall the definitions in Ref. \cite{two-photon coherence} for
the two length parameters $\Delta L$ and $\Delta L'$ and one phase
parameter $\Delta\phi$, which are given as:
\begin{eqnarray}\label{define}
\Delta L  &\equiv&  l_{1}-l_{2}= \left(\frac{l_{s1}+l_{i1}}{2}+l_{p1}\right)-\left(\frac{l_{s2}+l_{i2}}{2}+l_{p2}\right), \nonumber  \\
\Delta L' &\equiv&
l'_{1}-l'_{2}=\left(l_{s1}-l_{i1})-(l_{s2}-l_{i2}\right), \nonumber \\
\Delta\phi &\equiv&
\left(\phi_{s1}+\phi_{i1}+\phi_{p1}\right)-\left(\phi_{s2}+\phi_{i2}+\phi_{p2}\right).
\end{eqnarray}
Here the subscripts $p$, $s$ and $i$ stand for pump, signal and
idler respectively; $l$ denotes the optical path-length travelled
by a photon; and $\phi$ stands for phases other than the dynamic
one such as phase acquired due to reflections, geometric phase
etc. Thus, $l_{s1}$ denotes the path length travelled by the
signal photon in alternative $1$, etc.

The complete two-photon state $|\psi_{{\rm tp}}\rangle$ produced
by the double pass setup of Fig.~\ref{setup}(a) is the coherent
sum of the two-photon states produced in alternatives 1 and 2, and
using the definitions of Eq.~(\ref{define}) it can be written as
\begin{equation}
|\psi_{{\rm tp}}\rangle \ = \ |\psi_{{\rm tp}}\rangle_{1} \ + \
e^{-i(k_{0}\Delta L + \Delta\phi)} \ |\psi_{{\rm tp}}\rangle_{2}.
\end{equation}
Here $|\psi_{{\rm tp}}\rangle_{1(2)}$ is the two-photon state is
alternative $1(2)$; $k_{0}$ is the mean vacuum wavevector
magnitude of the pump wave and $k_{0}\Delta L + \Delta\phi$ is the
relative phase of the two-photon state in alternative 2 with
respect to the two-photon state in alternative 1. Interference is
observed in the coincidence count rate $R_{\textrm{AB}}$ of
detectors $D_{\textrm{A}}$ and $D_{\textrm{B}}$, which can be
calculated using the general expression derived in
Ref.\cite{two-photon coherence}:
\begin{eqnarray}\label{coin-count-rate1}
 R_{\textrm{AB}} = C \ [\  1+ \gamma(\Delta L)\gamma '(\Delta
L')
  \cos({k_{0}\Delta L+\Delta \phi})\ ].
\end{eqnarray}
Here C is a constant; $\gamma(\Delta L)$ is the pump correlation
function with a width equal to $l_{\textrm{coh}}^{p}$, the
coherence length of the pump, which in our case is about 5 cm;
$\gamma'(\Delta L')$ is the signal-idler correlation function with
a width equal to $l_{\textrm{coh}}$. The signal-idler coherence
length $l_{\textrm{coh}}$ is determined by the widths of the
interference filters and by the sizes of the iris diaphragms, and
is equal to about 100 $\mu$m in our case. From the two-photon path
diagram [Fig.~\ref{setup}($b$)] and using Eq.~(\ref{define}), we
find that $\Delta L=x_{s}+x_{i}$, $\Delta L'=2x_{s}-2x_{i}$ and
$\Delta\phi=2\beta_{s}+2\beta_{i}+\pi$. Assuming $\big{|}\Delta
L\big{|}\ll l_{\textrm{coh}}^{p}$ and $\big{|}\Delta L'\big{|}\ll
l_{\textrm{coh}}$, Eq.~(\ref{coin-count-rate1}) simplifies to
\begin{figure}[t!]
\includegraphics{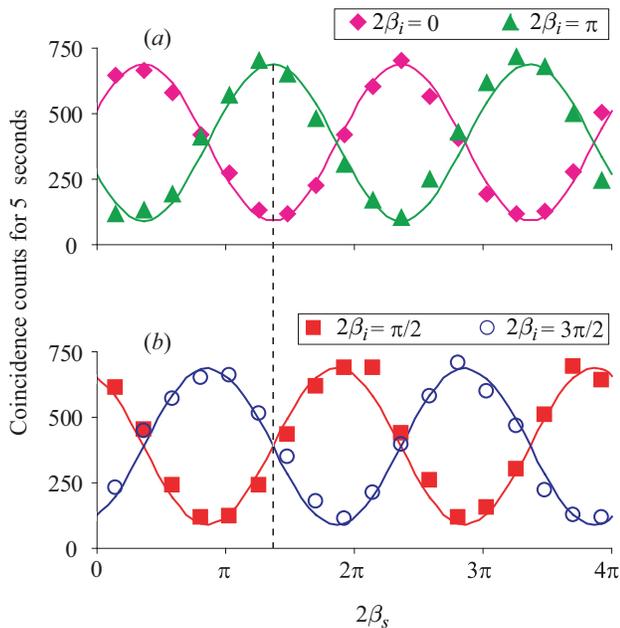}
  \caption{(color online). Measured number of coincidence counts as a function of 2$\beta_{s}$,
  the geometric phase of signal photon in alternative 1, for four different fixed values of 2$\beta_{i}$,
  the geometric phase of idler photon in alternative 1. ($a$) 2$\beta_{i}$=0 and $\pi$; ($b$) 2$\beta_{i}$=$\pi/2$ and $3\pi/2$.
  The solid lines are sinusoidal fits. The
distance of the dashed line from the origin is a measure of the
fixed value of $k_{0}(x_{s}+x_{i})$ to within the period
2$\pi$.}\label{experimental-result}
\end{figure}
\begin{equation}\label{count rate2}
R_{\textrm{AB}}= C \ \{ \  1-\cos[k_{0}(x_{s}+x_{i})
+2\beta_{s}+2\beta_{i}] \}.
\end{equation}
When the geometric phase $2\beta_s + 2\beta_i$ is  held fixed, the
variation of the coincidence rate $R_{\rm AB}$ with the dynamic
phase $k_0 (x_s + x_i)$ is of the form shown by Franson
\cite{bell-time-energy1} to lead to a violation of a CHSH-Bell
inequality. Bell inequality violations based on dynamic phase have
been reported in many experiments \cite{bell-time-energy2,
bell-time-energy3, hyper-polar+spatial+time-energy}. Similarly, we
note that when the dynamic phase $k_0 (x_s + x_i)$ is held fixed
and the geometric phase $2\beta_s + 2\beta_i$ is varied, the
nature of the variation of the coincidence rate $R_{\rm AB}$ is
still of the form to lead to a violation of the CHSH-Bell
inequality, but this time based solely on geometric phase.  We
next describe our experimental procedure for establishing a
violation of this inequality.

The experimental setup was initially aligned such that the
distances of the three mirrors from the crystal were all equal to
within a millimeter, and thus the condition $\big{|}\Delta
L\big{|}=\big{|}x_{s}+x_{i}\big{|}\ll l^{p}_{\textrm{coh}}$ was
satisfied. The idler mirror position was then scanned to observe
fringes in the coincidence count rate as a function of $x_{i}$,
and it was fixed at a position around which the observed fringe
visibility was maximum. At this position, $x_{s}$ and $x_{i}$ were
equal to within a few microns and thus the condition
$\big{|}\Delta L'\big{|}=\big{|}2x_{i}-2x_{s}\big{|}\ll
l_{\textrm{coh}}$ was adequately satisfied. Next, the quarter-wave
plate Q$_{\textrm{i2}}$ was successively fixed at four different
values of 2$\beta_{i}: 0, \pi/2, \pi, 3\pi/2$. For each value of
2$\beta_{i}$, coincidence counts were measured as a function of
2$\beta_{s}$. Fig.~\ref{experimental-result} shows the number of
coincidence counts plotted against $2\beta_s$ for four different
values of 2$\beta_{i}$.

The fringe visibilities shown in these plots are approximately
77\%. An experimental demonstration of a violation of a CHSH-Bell
inequality requires that the visibility of coincidence fringes be
greater than 70.7\% \cite{chsh, kwiat-high-intensity}. The value
of the Bell parameter $|\textrm{S}|$ is determined by the
visibility of the coincidence fringes \cite{bell-linear-momentum,
bell-time-energy3}. The sinusoidal coincidence fringes observed in
our experiment as functions of 2$\beta_{s}$  for four given values
of 2$\beta_{i}$ with visibilities of approximately 77\% imply that
the magnitude of the Bell parameter $|\textrm{S}|$ is
approximately 2.18 $\pm$ 0.04. Therefore, these measurements show
a violation of a Bell inequality by approximately 5 standard
deviations. A Bell inequality acts as an entanglement  witness,
and its violation verifies entanglement
\cite{hyper-polar+spatial+time-energy, entanglement witness}.
Thus, these results demonstrate that energy-time entanglement can
be explored using geometric phases of the signal and idler
photons. In our experiment, coincidences were collected for only 5
seconds. Our choice of the coincidence collection time was limited
by the overall stability of the interferometer. By collecting
coincidences for a longer period of time, a violation with an
increased number of standard deviations can be achieved.

Although 77\% visibility is sufficient to show a Bell inequality
violation, certain quantum information protocols require
visibilities closer to 100\%. The main reason for low visibility
in our experimental setup is the imperfect overlap of the two
interfering two-photon modes. This is caused by the relatively
large divergences of the signal and idler modes in alternative 1
compared to their divergences in alternative 2. This effect could
be minimized by using a single spherical mirror, with its center
of curvature located at the crystal, for reflecting the pump,
signal and idler modes \cite{cluster-mataloni-prl-07}. Another
reason for low visibility is the unequal coincidence count rates
in the two alternatives. This problem can be taken care of by
inserting a variable neutral density filter into the pump beam
path between the crystal and the pump mirror. The above factors
have also been noted to cause low visibilities in energy-time
entanglement experiments based on dynamic phase
\cite{bell-time-energy3}. We believe that by using customized
experimental setups, visibilities closer to the theoretical
maximum of 100\% should be achievable.

Geometric phase has found many applications in optics
\cite{hariharan-review, bhandari-review}. One of the
distinguishing features of geometric phase is its non-dispersive
nature. Dynamic phase is introduced by changing the optical path
length, and it remains inversely proportional to the wavelength.
However, geometric phase is a topological phase and does not
depend directly on the wavelength. The non-dispersive nature of
geometric phase has been demonstrated in white-light interference
using achromatic wave plates \cite{hariharan-jmo-94}.

For quantum information science, one potential benefit of using
geometric phase could be in exploring the energy-time entanglement
of ultrabroadband PDC sources \cite{broadband downconversion1,
broadband downconversion2}. For such sources, the signal-idler
coherence length $l_{\textrm{coh}}$ remains so small that the
visibility of two-photon fringes---which is the magnitude of the
product $\gamma(\Delta L)\gamma(\Delta L')$---does not remain
constant over the variations of $\Delta L'$ that are of the order
of a wavelength. Therefore, with such sources, dynamic phase is
unsuitable for performing Bell inequality violation experiments.
Geometric phase, on the other hand, is non-dispersive. Changing
geometric phase does not change the optical path lengths.
Therefore, the two-photon fringe visibility remains constant as a
function of the phase introduced geometrically, and this makes
geometric phase particularly suitable for exploring the
energy-time entanglement of ultrabroadband PDC sources.

Another benefit of using geometric phase lies in the ease of
introducing very small phase shifts. In contrast with dynamic
phase, where motorized translation stages are employed, geometric
phase is quite easily manipulated  by using rotating wave plates.
These features of geometric phase may also aid in the construction
of hyperentangled states \cite{hyper-polar+momentum1,
hyper-polar+momentum2, hyper-polar+spatial+time-energy} with
energy-time as one of the degrees of freedom.

In conclusion, we have observed a violation of the Bell inequality
for energy and time based purely on the geometric-phases of the
signal and idler photons produced by parametric down-conversion.
We have thus shown that energy-time entanglement can as well be
explored using the geometric phases of the signal and idler
photons. These results provide an additional means by which
entanglement can be manipulated, and therefore they may have
important practical implications for quantum information science.

We gratefully acknowledge financial support through a MURI grant
from the U.S. Army Research Office and through an STTR grant from
the U.S. Air Force Office. We thank John Sipe, Petros Zerom and
Heedeuk Shin for useful discussions.

\bibliographystyle{unsrt}

\end{document}